# Experimental Modal Analysis for engineering structures via time-delay Dynamic Mode Decomposition with Control

**Yanxin Si[1], Bayu Jayawardhana[1], J. Nathan Kutz[3], Yunpeng Zhu[2*], Liangliang Cheng[1*]**

[1] ENTEG, Faculty of Science and Engineering, University of Groningen, Groningen, The Netherlands;

[2] School of Engineering and Materials Science, Queen Mary University of London, London, UK;

[3] Department of Applied Mathematics, University of Washington, Seattle, USA

**Abstract**

Over the past decades, Experimental Modal Analysis (EMA) has been widely used to identify structural dynamic properties including natural frequency, damping ratio, and mode shape for assessing structural integrity. The most commonly adopted approach for EMA is the Poly reference Least Squares Complex Frequency (pLSCF) method, due to its strong capability to decouple closely spaced (dense) modes and its robustness to measurement noise. However, pLSCF-based EMA is typically applied to low-dimensional cases with a limited number of measurement points, because its computational cost increases rapidly for high-dimensional or continuous structural measurements, especially as the model order increases. To address this limitation, this paper develops a high-dimensional EMA framework based on Dynamic Mode Decomposition with control (DMDc) - a powerful data-driven technique originally applied in fluid dynamics - to advance modal identification under high-dimensional measurement scenarios. This study firstly clarifies the relationship between the pLSCF and time-delay DMDc using the discrete state-space representation of the auto-regressive with exogenous inputs (ARX) model of linear systems. By demonstrating that both methods reflect the same physical dynamics of the structure, it offers a physics-based rationale for applying time-delay DMDc in EMA. The capacity and advantages of using time-delay DMDc for modal parameters identification in both low and high dimensional measurements are validated via numerical simulations of a 6-degree of freedom (6-DoF) system and experiments on a cantilever beam using digital camera, respectively. This study demonstrates the effectiveness of the time-delay DMDc in robust and reliable modal parameter identification, successfully addressing high-dimensional EMA challenges that conventional pLSCF cannot resolve, and highlighting its potential for real-world structural dynamics applications.

## 1 Introduction

Experimental Modal Analysis (EMA) is a well-established methodology for identifying the modal characteristics of structures, including natural frequencies, damping ratios, and mode shapes[1]. In EMA, the structure is often subjected to free vibration or with controlled excitations, such as impact, swept-sine, or chirp inputs. The resulting responses, including acceleration, velocity, or displacement, are measured for subsequent modal parameter estimation. In practice, EMA plays a critical role in structural health monitoring by providing high-fidelity modal identification under controlled conditions. This accuracy is essential across a wide range of engineering applications, including aerospace structures[2], civil infrastructure[3, 4], and mechanical systems[5].

Basically, EMA can be implemented in either the time domain or the frequency domain. Time domain approaches, such as the Ibrahim Time Domain (ITD) method[6], ARX models[7], and Stochastic Subspace Identification (SSI)[8], estimate modal parameters directly by fitting parametric dynamic models (e.g., exponential decay functions, autoregressive representations, or state-space models) based on measured input–output time histories. However, these methods are less frequently adopted in practical EMA due to their inherent sensitivity to measurement and modeling noise. Such sensitivity often leads to unstable modal parameter estimates, particularly for damping ratios[9], as well as the computational challenges associated with full-field, high-dimensional measurements acquired from large-scale systems using techniques like digital cameras or scanning laser Doppler vibrometers[10].

Frequency domain methods such as the Frequency Domain Decomposition (FDD)[11], Complex Mode Indicator Function (CMIF)[12], and poly-reference Least Squares Complex Frequency-domain (pLSCF) methods[13], are more commonly adopted in practice than time domain methods, due to their enhanced robustness to measurement noise and their ability to clearly identify closely spaced modes through stabilization diagrams. Among these methods, the pLSCF approach is an industrial benchmark for EMA through the construction of a poly-reference rational model of the measured Frequency Response Functions (FRFs) using a Least-Squares (LS) error-minimization formulation[14]. The modal parameters are subsequently extracted by the poles/zeros of the fitted FRFs[15]. However, the computational cost and memory requirements of pLSCF increase rapidly with data dimensionality, particularly when higher model orders are required to ensure a stable estimation of damping ratios. For example, Cunha and Caetano noted that identifying large-scale structures frequently requires excessively high model orders to ensure damping ratio convergence. This not only increases the computational burden, but also causes a number of spurious mathematical poles, which complicates the extraction of true physical modal parameters[16]. This significantly limits its applicability to full-field, high-dimensional measurements in large-scale systems.

Although recent advances in machine learning-based EMA, such as Convolutional Neural Networks (CNNs)[17], Physics-Informed Neural Networks (PINNs)[18], and Bayesian deep learning[19], have been developed to automatically identify modal parameters from high-dimensional data, they are still limited in physical interpretability and generalization capabilities. For instance, a model trained on a bridge under specific thermal conditions may fail to generalize when environmental temperatures shift, as it relies on statistical correlations within the training data rather than the fundamental physical laws of structural dynamics[20]. Therefore, there is an urgent need for alternative approaches that can effectively handle high-dimensional data while maintaining robustness to measurement error.

Dynamic Mode Decomposition (DMD) approaches, such as DMDc[21], Multi-resolution DMD[22], Koopman-DMD[23], were originally developed for fluid dynamics as a purely data-driven dimensionality-reduction method to extract coherent spatiotemporal structures from high-dimensional flow fields. In fluid dynamics, DMD evaluates eigenvalues and eigenvectors that directly encode the characteristic frequencies, decay rates (damping), and coherent spatial mode structures of the flow. For instance, Schmid demonstrated that DMD can extract dynamic modes from complex flow sequences, such as wakes and jets, without a prior numerical model[24]. By projecting large-scale datasets onto a low-dimensional subspace, the method effectively isolates localized instabilities while accurately capturing characteristic frequencies and decay rates. It is therefore natural to consider extending DMD-based approaches to structural EMA. Recently, a few studies have strived to apply DMD approaches to structure engineering for high-dimensional EMA. For example, Saito et al. demonstrated the applicability of DMD to EMA of linear mechanical systems under free vibration, showing accurate modal parameter estimation for high-dimensional data under low measurement noise conditions. However, it failed to identify stable damping ratios using high measurement noise vibration data[25].

The limitations of standard DMD can be effectively mitigated by incorporating time-delay embedding, as seen in approaches such as higher-order dynamic mode decomposition (HODMD)[26] and time-delay DMD[27]. These approaches use temporal correlations to improve robustness to measurement noise and stabilize the estimation of DMD eigenvalues. Thus, these advanced variants offer superior stability in damping ratio identification. For instance, Cheng et al. demonstrated the high potential of combining the CoTracker model with time-delay DMD to extract full-field modal parameters, including natural frequencies, damping ratios and mode shapes, from video sequences[28].

However, existing DMD-based EMA studies have been largely limited to free-vibration scenarios, leaving three challenging research questions unresolved as below:

1) What fundamental principles enable DMD-based approaches, originally developed for fluid dynamics, to be applicable to the analysis of structural dynamics?
2) How to apply the DMD-based approaches for a systematic EMA of structures using high-dimensional measurement data under prescribed input excitations?
3) How to achieve robust and stable estimation of modal parameters, particularly damping ratios, in high-dimensional EMA?

In order to answer these questions, the present study introduces a time-delay DMDc-based EMA method. Here, the primary novelties are summarized as follows:
1) DMDc has been widely used in fluid dynamics, but the physical interpretability of its data-driven modes and their applicability to structural dynamics remain unclear. This study forms a theoretical basis for extending DMDc to structural modal analysis.
2) This study integrates time-delay states into DMDc to enable robust high-dimensional EMA, a challenge that existing modal analysis methods (e.g., pLSCF and CMIF) cannot readily address.
3) The proposed approach is also effective for low-dimensional EMA, delivering comparable mode identification while providing more stable damping estimates than pLSCF.
4) Representative case studies were conducted on both low-dimensional (6-DoF) under severe noise conditions (signal-to-noise ratio (SNR) down to 10 dB) and high-dimensional (beam vibration video) structures, demonstrating the practical effectiveness of the proposed method as a new paradigm for structural EMA.

The remainder of this paper is organized as follows. Section 2 presents the theoretical foundations of the proposed time-delay DMDc method. Section 3 evaluates its performance on a simulated 6-DoF structure and compares it against the pLSCF. Section 4 applies the method to cantilever beam experiments using both accelerometer and full-field video measurements under hammer excitation. Finally, Section 5 summarizes the main findings and discusses the broader implications of time-delay DMDc for future developments in EMA.

## 2. Time-delay DMDc for high-dimensional dynamic systems

### 2.1 Relationship between time-delay DMDc and pLSCF

Uncovering the theoretical relationship between DMDc approaches and the pLSCF method provides a fundamental basis for establishing the reliability and physical fidelity of DMDc-based EMA of high-dimensional measurement data. The following analysis demonstrates that the modal parameters of a structure obtained using time-delay DMDc are mathematically equivalent to those identified using the pLSCF method. This equivalence is rigorously proven by formulating both approaches within a standard discrete-time ARX framework, which serves as a unifying bridge between time-delay DMDc and pLSCF.

*1) pLSCF and ARX model*

Considering a linear time invariant system that has $n$ degree of freedom with $m$ external excitations, where $n, m \in \mathbb{Z}^+$, the input excitations and output responses of the system is denoted as

$$\mathbf{u}_k \underset{m \times 1}{=} \begin{bmatrix} u_1(k) \\ \vdots \\ u_m(k) \end{bmatrix} \text{ and } \mathbf{y}_k \underset{n \times 1}{=} \begin{bmatrix} y_1(k) \\ \vdots \\ y_n(k) \end{bmatrix} \quad (1)$$

where $k = 1, \ldots, K$ represents the discrete time with the sampling time of $\Delta t$.

Taking the Discrete-Time Fourier Transform (DTFT) of $\mathbf{y}_k$ and $\mathbf{u}_k$, the corresponding spectra are achieved as

$$\mathbf{U}_\omega = \begin{bmatrix} U_1(j\omega) \\ \vdots \\ U_m(j\omega) \end{bmatrix}_{m \times 1} \text{ and } \mathbf{Y}_\omega = \begin{bmatrix} Y_1(j\omega) \\ \vdots \\ Y_n(j\omega) \end{bmatrix}_{n \times 1} \quad (2)$$

where $\omega$ is the frequency (rad/s), such that the system frequency domain representation is

$$\mathbf{Y}_\omega = \mathbf{H}_\omega \mathbf{U}_\omega \quad (3)$$

Where $\mathbf{H}_\omega \in \mathbb{C}^{n \times m}$ is the FRF matrix.

In the pLSCF method, the system frequency response function (FRF) matrix is represented as a rational function sharing a common denominator polynomial, such that

$$\mathbf{H}_{\omega,\text{pLSCF}} = \mathbf{N}_\omega D^{-1}(j\omega) \quad (4)$$

where $\mathbf{N}_\omega \in \mathbb{C}^{n \times m}$ is the numerator matrix, and $D(j\omega) = d_0 + d_1 q^{-1} + \cdots + d_\tau q^{-\tau}$ with $\tau \in \mathbb{Z}^+$ and $q = e^{j\omega \Delta t}$ is the common denominator polynomial.

On the other hand, from an ARX model of the system

$$\mathbf{y}_k + \mathbf{a}_1 \mathbf{y}_{k-1} + \cdots + \mathbf{a}_{\tau_a} \mathbf{y}_{k-\tau_a} = \mathbf{b}_1 \mathbf{u}_{k-1} + \cdots + \mathbf{b}_{\tau_b} \mathbf{u}_{k-\tau_b} \quad (5)$$

where $\mathbf{a}_1, \ldots, \mathbf{a}_{\tau_a} \in \mathbb{R}^{n \times n}$ and $\mathbf{b}_1, \ldots, \mathbf{b}_{\tau_b} \in \mathbb{R}^{n \times m}$ are the coefficient matrices with $\tau_a, \tau_b \in \mathbb{Z}^+$, the FRF matrix can be evaluated as

$$\mathbf{H}_{\omega,\text{ARX}} = \mathbf{A}_\omega^{-1} \mathbf{B}_\omega \quad (6)$$

where $\mathbf{A}_\omega = \mathbf{I} + \mathbf{a}_1 q^{-1} + \cdots + \mathbf{a}_{\tau_a} q^{-\tau_a}$ and $\mathbf{B}_\omega = \mathbf{b}_1 q^{-1} + \cdots + \mathbf{b}_{\tau_b} q^{-\tau_b}$, $\mathbf{I}$ is the $n \times n$ unit matrix.

Considering

$$\mathbf{A}_\omega^{-1} = \frac{\text{adj}(\mathbf{A}_\omega)}{\det(\mathbf{A}_\omega)} \quad (7)$$

where $\text{adj}(\mathbf{A}_\omega)$ is an adjugate matrix whose elements are polynomials in $e^{j\omega \Delta t}$, while $\det(\mathbf{A}_\omega)$ is the determinant, a scalar polynomial in $e^{j\omega \Delta t}$. In this case, let $\mathbf{N}_\omega = \text{adj}(\mathbf{A}_\omega) \mathbf{B}_\omega$ and $D(j\omega) = \det(\mathbf{A}_\omega)$, the FRF matrix of an ARX model is equivalent to that from pLSCF.

*2) ARX model and time-delay DMDc*

ARX model can be written as the discrete state-space equation. The discretized state-space representation of Eq. (5) is derived as follows. Let $\mathbf{x}_{0,k}, \ldots, \mathbf{x}_{\tau_a,k} \in \mathbb{R}^n$ as a set of state vectors for the system:

$$\begin{cases} \mathbf{x}_{0,k} = \mathbf{y}_k \\ \quad \vdots \\ \mathbf{x}_{\tau_a,k} = \mathbf{y}_{k-\tau_a} \end{cases} \quad (8)$$

These state variables are then related by:

$$\begin{cases} \mathbf{x}_{\tau_a,k} = \mathbf{x}_{\tau_a-1,k-1} \\ \quad \vdots \\ \mathbf{x}_{0,k} = -\mathbf{a}_1 \mathbf{x}_{0,k-1} - \ldots - \mathbf{a}_{\tau_a} \mathbf{x}_{\tau_a-1,k-1} + \mathbf{b}_1 \mathbf{u}_{k-1} + \ldots + \mathbf{b}_{\tau_b} \mathbf{u}_{k-\tau_b} \end{cases} \quad (9)$$

Writing the above equation in matrix-vector form yields the discrete state-space equation:

$$\bar{\mathbf{x}}_k = \mathbf{A}_{ARX}\bar{\mathbf{x}}_{k-1} + \mathbf{B}_{ARX}\bar{\mathbf{u}}_{k-1} \tag{10}$$

$$\mathbf{y}_k = \mathbf{C}\bar{\mathbf{x}}_k \tag{11}$$

The two state and input vectors in Eq. (10) are expressed below:

$$\underset{n(\tau_a+1)\times 1}{\bar{\mathbf{x}}_k} = \begin{bmatrix} \mathbf{x}_{0,k} \\ \vdots \\ \mathbf{x}_{\tau_a,k} \end{bmatrix},\quad \underset{n(\tau_a+1)\times 1}{\bar{\mathbf{x}}_{k-1}} = \begin{bmatrix} \mathbf{x}_{0,k-1} \\ \vdots \\ \mathbf{x}_{\tau_a,k-1} \end{bmatrix} \text{ and } \underset{m\tau_b\times 1}{\bar{\mathbf{u}}_{k-1}} = \begin{bmatrix} \mathbf{u}_{k-1} \\ \vdots \\ \mathbf{u}_{k-\tau_b} \end{bmatrix} \tag{12}$$

Where the state matrix $\mathbf{A}_{ARX}$, input matrix $\mathbf{B}_{ARX}$ and output matrix $\mathbf{C}$ are given as:

$$\underset{(\tau_a+1)n\times(\tau_a+1)n}{A_{ARX}} = \begin{bmatrix} -\mathbf{a}_1 & -\mathbf{a}_2 & \cdots & -\mathbf{a}_{\tau_a} & 0 \\ \mathbf{I} & 0 & \cdots & 0 & 0 \\ 0 & \mathbf{I} & \cdots & 0 & 0 \\ \vdots & \vdots & \ddots & \vdots & \vdots \\ 0 & 0 & \cdots & \mathbf{I} & 0 \end{bmatrix} \text{ and } \underset{(\tau_a+1)n\times(\tau_b m)}{\mathbf{B}_{ARX}} = \begin{bmatrix} \mathbf{b}_1 & \mathbf{b}_2 & \cdots & \mathbf{b}_{\tau_b-1} & \mathbf{b}_{\tau_b} \\ 0 & 0 & \cdots & 0 & 0 \\ \vdots & \vdots & \ddots & \vdots & \vdots \\ 0 & 0 & \cdots & 0 & 0 \\ 0 & 0 & \cdots & 0 & 0 \end{bmatrix} \tag{13}$$

$$\underset{n\times(\tau_a+1)n}{\mathbf{C}} = \begin{bmatrix} \mathbf{I} & 0 & \cdots & 0 \end{bmatrix} \tag{14}$$

Where $\mathbf{I}$ is the unit $n\times n$ matrix.

Inspired by the use of time-delay states in the state-space representation of the ARX model, time-delay DMDc emerges as an extension of the standard DMDc, which combines the ideas of standard DMDc and Takens' time-delay theorem to augment a more versatile and robust state matrix represented by a few measurements with their former time-lagged snapshots[29]. It is a classic technique for dealing with long-term temporal behavior and high-dimensional data within the areas of vibration control and system parameter identification.

To implement the time-delay DMDc, let $\tau_a$ and $\tau_b$ denote the time-delay orders for the output and input. Stacking state vectors $\tau_a \le K$ to form the following augmented matrix:

$$\underset{(\tau_a+1)n\times K}{\mathbf{D}} = \begin{bmatrix} \mathbf{x}_{0,1} & \mathbf{x}_{0,2} & \cdots & \mathbf{x}_{0,K} \\ \mathbf{x}_{1,1} & \mathbf{x}_{1,2} & \cdots & \mathbf{x}_{1,K} \\ \vdots & \vdots & \ddots & \vdots \\ \mathbf{x}_{\tau_a,1} & \mathbf{x}_{\tau_a,2} & \cdots & \mathbf{x}_{\tau_a,K} \end{bmatrix} = \begin{bmatrix} \bar{\mathbf{x}}_1 & \cdots & \bar{\mathbf{x}}_K \end{bmatrix} \tag{15}$$

Consider a snapshot matrix $\mathbf{X}$ containing snapshots of the generalized coordinates $\bar{\mathbf{x}}_k, k=1,\ldots,K-1$ as its column vectors, furthermore, define a time shifted snapshot matrix $\mathbf{X}'$ containing snapshots of the generalized coordinates $\bar{\mathbf{x}}_k, k=2,\ldots,K$ as its column vectors and the current control matrix $\boldsymbol{\Gamma}$.

$$\underset{n(\tau_a+1)\times(K-1)}{\mathbf{X}} = \begin{bmatrix} \bar{\mathbf{x}}_1 & \bar{\mathbf{x}}_2 & \cdots & \bar{\mathbf{x}}_{K-1} \end{bmatrix} \text{ and } \underset{n(\tau_a+1)\times(K-1)}{\mathbf{X}'} = \begin{bmatrix} \bar{\mathbf{x}}_2 & \bar{\mathbf{x}}_3 & \cdots & \bar{\mathbf{x}}_K \end{bmatrix} \tag{16}$$

$$\underset{(m\tau_b)\times(K-1)}{\boldsymbol{\Gamma}} = \begin{bmatrix} \bar{\mathbf{u}}_1 & \bar{\mathbf{u}}_2 & \cdots & \bar{\mathbf{u}}_{K-1} \end{bmatrix} \tag{17}$$

By reformulating Eq. (5) using the newly defined matrices $\mathbf{X}', \mathbf{X}$ and $\boldsymbol{\Gamma}$, the time-delay DMDc formulation can be expressed as:

$$\mathbf{X}' = \mathbf{A}_{DMDc}\mathbf{X} + \mathbf{B}_{DMDc}\boldsymbol{\Gamma} \tag{18}$$

The linear dynamical system connecting the future state relies on information from both the current state and the current control. Comparing Eq. (5) with Eq. (18) demonstrates that the state matrix $\mathbf{A}_{\text{DMDc}}$ in time-delay DMDc is identical to the ARX state matrix $\mathbf{A}_{\text{ARX}}$. Therefore, all three methods yield identical system eigenvalues and modal parameters, which are independent of the excitation amplitude as demonstrated in the following proposition.

**Proposition 1:** The eigenvalues identified using time-delay DMDc are independent of the excitation amplitude.

**Proof 1:** Scaling the inputs by a set of nonzero constant factor $\alpha_1, \ldots, \alpha_m$:

$$u_i'(k) = \alpha_i u_i(k), i = 1, \ldots, m \tag{19}$$

Define the scaling matrix:

$$\mathbf{u}_k' = \mathbf{\Lambda} \mathbf{u}_k, \mathbf{\Lambda} = \text{diag}(\alpha_1, \alpha_2, \ldots, \alpha_m) \tag{20}$$

The input vector and the current control matrix are:

$$\bar{\mathbf{u}}_k' = (\mathbf{I}_{\tau_b} \otimes \mathbf{\Lambda}) \bar{\mathbf{u}}_k \tag{21}$$

$$\bar{\mathbf{\Gamma}} = (\mathbf{I}_{\tau_b} \otimes \mathbf{\Lambda}) \mathbf{\Gamma} \tag{22}$$

Given the system is LTI system, the scaling of input inherently leads to proportional scaling of outputs.

$$\mathbf{y}_k' = \mathbf{P} \mathbf{y}_k \tag{23}$$

Thus, the delay state satisfies:

$$\bar{\mathbf{x}}_k' = (\mathbf{I}_{\tau_a+1} \otimes \mathbf{P}) \bar{\mathbf{x}}_k, \bar{\mathbf{X}}' = (\mathbf{I}_{\tau_a+1} \otimes \mathbf{P}) \mathbf{X} \text{ and } \bar{\mathbf{X}} = (\mathbf{I}_{\tau_a+1} \otimes \mathbf{P}) \mathbf{X} \tag{24}$$

The DMDc is therefore rewritten as

$$(\mathbf{I}_{\tau_a+1} \otimes \mathbf{P}) \mathbf{X}' = \mathbf{A}_{\text{DMDc}} (\mathbf{I}_{\tau_a+1} \otimes \mathbf{P}) \mathbf{X} + \mathbf{B}_{\text{DMDc}} (\mathbf{I}_{\tau_b} \otimes \mathbf{\Lambda}) \mathbf{\Gamma} \tag{25}$$

Left-multiply by $\left(\mathbf{I}_{\tau_a+1} \otimes \mathbf{P}\right)^{-1}$:

$$\mathbf{X}' = \mathbf{A}_{\text{DMDc}} \mathbf{X} + \underbrace{\left(\mathbf{I}_{\tau_a+1} \otimes \mathbf{P}\right)^{-1} \mathbf{B}_{\text{DMDc}} \left(\mathbf{I}_{\tau_b} \otimes \mathbf{\Lambda}\right)}_{\bar{\mathbf{B}}_{\text{DMDc}}} \tag{26}$$

It can be seen that scaling the input amplitude by a constant factor only rescales the identified input matrix $\mathbf{B}_{\text{DMDc}}$, while the state matrix $\mathbf{A}_{\text{DMDc}}$ remains unchanged. The resulting modal parameters, defined by eigenvalues and eigenvectors of $\mathbf{A}_{\text{DMDc}}$, are independent of the amplitude of the excitation. While the mathematical equivalence between the time-delay DMDc and the ARX model is well-established, DMDc provides distinct advantages for high-dimensional EMA comparing ARX model[30].

1) The ARX model is a parametric approach that requires the explicit estimation of polynomial coefficients. Its performance is dependent on the precise selection of the time-delay order, especially in the presence of measurement noise. In contrast, the DMDc approach is model-free framework, seeking the optimal linear operator to characterize state evolution. It makes DMDc exceptionally robust for analyzing complex modern structures, particularly when the underlying physical mechanisms are partially unknown or too complicated for conventional parametric identification.

2) For high-dimensional EMA, the ARX coefficient matrices become computationally prohibitive to solve via standard least-squares regression. Conversely, DMDc integrates Singular Value Decomposition (SVD) to project high-dimensional snapshots into a low-dimensional subspace. This allows for the identification of the system's dominant dynamics without the need to solve for the full-order ARX parameters.

**2.2 Modal parameter identification using time-delay DMDc**

The results presented in Section 2.1 demonstrate that the modal parameters identified using time-delay DMDc are mathematically equivalent to those obtained via the pLSCF method. For high-dimensional systems, time-delay DMDc enables efficient estimation of system eigenvalues and eigenvectors in a reduced-order state space, thereby facilitating reliable modal parameter identification as discussed in follows.

In the time-delay DMDc model in Eq. (18), conducting the SVD of the output matrix $\mathbf{X}' = \mathbf{U}\mathbf{\Sigma}\mathbf{V}^* \approx \hat{\mathbf{U}}\hat{\mathbf{\Sigma}}\hat{\mathbf{V}}^*$, where $\hat{\mathbf{U}}$ is the truncated matrix of the left unitary matrix $\mathbf{U}$ with the truncation order $r$, $\hat{\mathbf{V}}$ is the truncated matrix of the right unitary matrix $\mathbf{V}$, $\hat{\mathbf{\Sigma}}$ is the truncated diagonal matrix of the eigenvalue matrix $\mathbf{\Sigma}$. Here "*" denotes the complex conjugate transpose. Denote the reduced-order state is $\mathbf{X} = \hat{\mathbf{U}}\bar{\mathbf{X}}$, the time delay DMDc can be obtained as

$$\bar{\mathbf{X}}' = \hat{\mathbf{U}}^*\mathbf{A}_{\mathrm{DMDc}}\hat{\mathbf{U}}\bar{\mathbf{X}} + \hat{\mathbf{U}}^*\mathbf{B}_{\mathrm{DMDc}}\mathbf{\Gamma} \tag{27}$$

or

$$\bar{\mathbf{X}}' = \begin{bmatrix} \hat{\mathbf{U}}^*\mathbf{A}_{\mathrm{DMDc}} & \hat{\mathbf{U}}^*\mathbf{B}_{\mathrm{DMDc}} \end{bmatrix} \begin{bmatrix} \mathbf{X} \\ \mathbf{\Gamma} \end{bmatrix} = \hat{\mathbf{G}}\mathbf{\Omega} \tag{28}$$

Considering the truncated SVD of the input matrix $\mathbf{\Omega} \approx \tilde{\mathbf{U}}\tilde{\mathbf{\Sigma}}\tilde{\mathbf{V}}^*$ with truncation order $p$, there is[21]

$$\hat{\mathbf{G}} = \bar{\mathbf{X}}'\tilde{\mathbf{V}}\tilde{\mathbf{\Sigma}}^{-1}\tilde{\mathbf{U}}^* \tag{29}$$

The matrix $\tilde{\mathbf{U}}$ is spilt into two matrices, $\tilde{\mathbf{U}} = \begin{bmatrix} \tilde{\mathbf{U}}_1^* & \tilde{\mathbf{U}}_2^* \end{bmatrix}^*$ to provide bases for $\mathbf{X}$ and $\mathbf{\Gamma}$, such that

$$\begin{cases} \hat{\mathbf{U}}^*\mathbf{A}_{\mathrm{DMDc}} = \bar{\mathbf{X}}'\tilde{\mathbf{V}}\tilde{\mathbf{\Sigma}}^{-1}\tilde{\mathbf{U}}_1^* = \hat{\mathbf{U}}^*\mathbf{X}'\tilde{\mathbf{V}}\tilde{\mathbf{\Sigma}}^{-1}\tilde{\mathbf{U}}_1^* \\ \hat{\mathbf{U}}^*\mathbf{B}_{\mathrm{DMDc}} = \bar{\mathbf{X}}'\tilde{\mathbf{V}}\tilde{\mathbf{\Sigma}}^{-1}\tilde{\mathbf{U}}_2^* = \hat{\mathbf{U}}^*\mathbf{X}'\tilde{\mathbf{V}}\tilde{\mathbf{\Sigma}}^{-1}\tilde{\mathbf{U}}_2^* \end{cases} \tag{30}$$

The reduced linear operator of $\bar{\mathbf{X}}$ in (21) is therefore

$$\tilde{\mathbf{A}}_{\mathrm{DMDc}} = \hat{\mathbf{U}}^*\mathbf{A}_{\mathrm{DMDc}}\hat{\mathbf{U}} = \hat{\mathbf{U}}^*\mathbf{X}'\tilde{\mathbf{V}}\tilde{\mathbf{\Sigma}}^{-1}\tilde{\mathbf{U}}_1^*\hat{\mathbf{U}} \tag{31}$$

From the studies of DMD approaches[30], the eigenvalues of $\tilde{\mathbf{A}}_{\mathrm{DMDc}}$ is the same as the first $r$ dominant eigenvalues of $\mathbf{A}_{\mathrm{DMDc}}$. The eigenvectors $\mathbf{\Phi} = [\boldsymbol{\phi}_1, \ldots, \boldsymbol{\phi}_r]$ of the original operator $\mathbf{A}_{\mathrm{DMDc}}$ is reconstructed by using the exact DMD method[31], representing the exact modal shapes

$$\mathbf{\Phi} = \mathbf{X}'\tilde{\mathbf{V}}\tilde{\mathbf{\Sigma}}^{-1}\tilde{\mathbf{U}}_1^*\hat{\mathbf{U}}\mathbf{W} \tag{32}$$

where $\tilde{\mathbf{A}}_{\mathrm{DMDc}}\mathbf{W} = \mathbf{W}\mathbf{\Lambda}$, $\mathbf{W} = [\mathbf{w}_1, \ldots, \mathbf{w}_r]$ and $\mathbf{\Lambda} = \mathrm{diag}(\mu_1, \ldots, \mu_r)$ are the eigenvectors and eigenvalues, respectively.

To extract natural frequency and damping ratio, discrete-time eigenvalues $\mu_i (i = 1, \ldots, r)$ are mapped to continuous-time eigenvalues $s_i$.

$$s_i = \frac{\log(\mu_i)}{\Delta t} = -\zeta_i\omega_i \pm j\omega_i\sqrt{1-\zeta_i^2} \tag{33}$$

Natural frequency $f_i$ and damping ratio $\zeta_i$ are derived from the eigenvalues $s_i$, $\omega_i$ denotes the natural angular frequency.

$$f_i = \frac{\omega_i}{2\pi} \quad (34)$$

$$\zeta_i = \frac{-\mathrm{Re}(s_i)}{|s_i|} \quad (35)$$

In summary, time-delay DMDc has proven to be particularly effective in handling high-dimensional data and can decompose complex systems into a set of spatially coherent modes, each associated with a specific frequency and growth/decay rate. In the following sections, its effectiveness is validated using simulated data from a 6-DoF system and the beam structure, and the results are compared against those obtained from the pLSCF method.

### 2.3 Time-delay order selection for the robustness of DMDc-based EMA

From the above discussion, it is evident that the modal parameters of high-dimensional systems can be accurately identified using time-delay DMDc within a reduced-order state space. However, the choice of time-delay order has a significant impact on the accuracy and robustness of time-delay DMDc-based EMA. In the following, the selection of the time-delay order under different excitation conditions is investigated, leading to the construction of frequency- and damping-ratio pseudo-stabilization diagrams.

For the augmented matrix **D**, each row corresponds to the time series of the signal with a specific delay $d \in \{0,\ldots,\tau_a\}$, producing $\tau_a + 1$ delayed sequences. Each column of the matrix constitutes a single time-delay embedding window, formed by stacking these delayed signal segments.

In the ideal, noise-free case, a minimal embedding with $\tau_a = 1$ is sufficient.

In practice, measurement noise motivates a larger delay order such that the embedding window spans at least one period of the lowest-frequency mode $f_{\min}$ this practical requirement can be expressed as

$$(\tau_a + 1)\Delta t \geq \frac{1}{f_{\min}} \quad \Rightarrow \quad \tau_a \geq \frac{f_s}{f_{\min}} - 1 \quad (36)$$

Increasing $\tau_a$ generally improves robustness to measurement noise, particularly for damping ratio estimation (see Appendix A). For impulse tests, it is recommended to form the delay-embedding snapshots from the portion of the ring-down with adequate SNR (i.e., avoid including very late, noise-dominated samples).

For special excitation types other than transient impacts, swept-sine or chirp excitation (nonstationary excitation) produces a response with a time-varying instantaneous frequency. Each delay time series represents a different temporal segment of this response, Thus, to ensure that all system modes are adequately captured within every delayed window, the following steps are required.

(a) Preprocessing via zero-padding

When the acquired data length is insufficient to cover the full system modes in every delayed window, zero-padding of $\tau_a$ and $\tau_b$ samples need to be applied to both ends of the excitation and response sequences respectively, as established in Eq. (37). The construction of the state vectors proceeds using the newly defined $\tilde{\mathbf{y}}_k$ and $\tilde{\mathbf{u}}_k$.

$$\tilde{\mathbf{y}}_k = \begin{cases} 0, & k < \tau_a + 1 \text{ or } k > K + \tau_a, \\ \mathbf{y}_k, & 1 \leq k \leq K, \end{cases} \quad \tilde{\mathbf{u}}_k = \begin{cases} 0, & k < 1 + \tau_b \text{ or } k > K + \tau_b, \\ \mathbf{u}_k, & 1 \leq k \leq K. \end{cases} \tag{37}$$

It effectively eliminates boundary spectral leakage and preserves the uniform frequency representation required for accurate modal parameter identification.

(b) Delay order bounds

For delay order $\tau_a$ and total sample count $K$, the embedding generates $\tau_a + 1$ time-shifted windows, where $i^{\text{th}}$ window ($i = 0, 2, \ldots, \tau_a$) spans samples $k \in [i, K - \tau_a + i]$. To guarantee that every window captures the complete sweep range from the lowest to the highest natural frequency, two boundary conditions must be satisfied: (1) Start-point constraint: The last window ($i = \tau_a$) must begin no later than the sample point $k_{\min}$ where the sweep reaches the lowest natural frequency; (2) End-point constraint: The first window ($i = 1$) must end no earlier than the sample point $k_{\max}$ where the sweep reaches the highest natural frequency. Combining these conditions yields the practical upper bound for the delay order:

$$\tau \leq \min(k_{\min}, K - k_{\max}) \tag{38}$$

## 3. Numerical examples of six degree-of-freedom system

### 3.1 EMA without measurement noises

Consider a representative 6-DoF storey building system shown in Fig.1 (a). The equations of motion can be written as a matrix form:

$$\mathbf{M}\ddot{\mathbf{y}}(t) + \mathbf{C}\dot{\mathbf{y}}(t) + \mathbf{K}\mathbf{y}(t) = \mathbf{u}(t) \tag{39}$$

where $\mathbf{y}(t) = [y_1(t), \ldots, y_6(t)]^{\text{T}}$ is the response vector, $\mathbf{u}(t) = [u(t), 0, \ldots, 0]^{\text{T}}$ is the input vector. $\mathbf{M}, \mathbf{C}$ and $\mathbf{K}$ are mass, damping and stiffness matrices, respectively, defined as

$$\mathbf{M} = \begin{pmatrix} m & 0 & 0 & 0 & 0 & 0 \\ 0 & m & 0 & 0 & 0 & 0 \\ 0 & 0 & m & 0 & 0 & 0 \\ 0 & 0 & 0 & m & 0 & 0 \\ 0 & 0 & 0 & 0 & m & 0 \\ 0 & 0 & 0 & 0 & 0 & m \end{pmatrix} \tag{40}$$

$$\mathbf{K} = \begin{bmatrix} 2k & -k & 0 & 0 & 0 & 0 \\ -k & 2k & -k & 0 & 0 & 0 \\ 0 & -k & 2k & -k & 0 & 0 \\ 0 & 0 & -k & 2k & -k & 0 \\ 0 & 0 & 0 & -k & 2k & -k \\ 0 & 0 & 0 & 0 & -k & k \end{bmatrix} \tag{41}$$

with m = 1 kg, k = 4 N·s/m, and $\mathbf{C} = 0.02\mathbf{M} + 0.0001\mathbf{K}$.

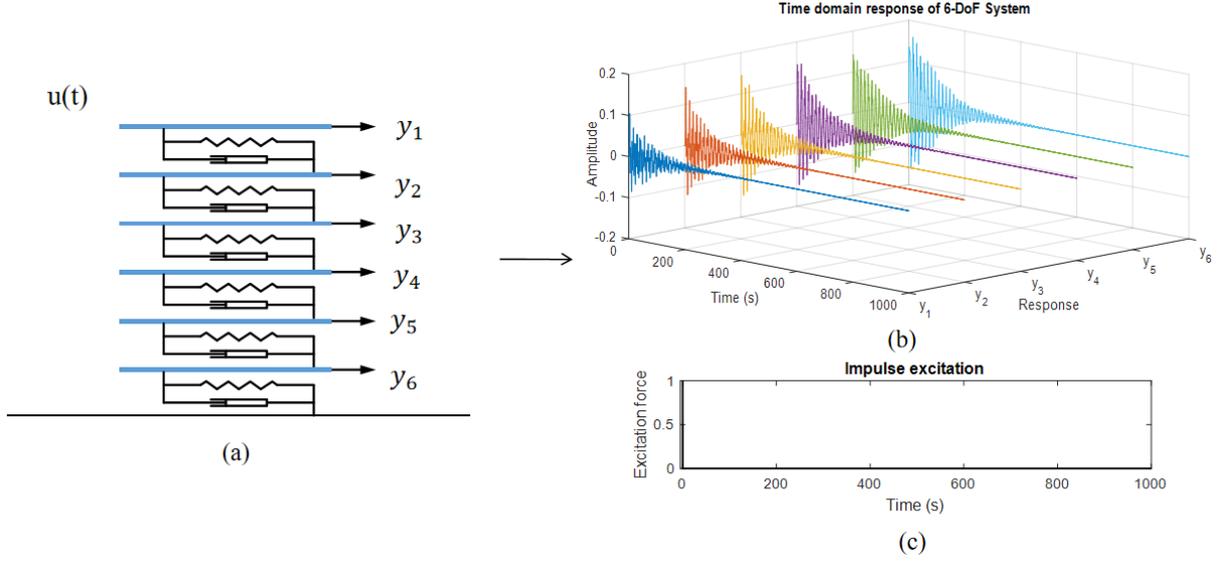

Fig. 1. (a) Configuration of a 6-DOF spring-mass-damper system, (b) An impulsive input and (c) Its time-history responses.

By using Eigenvalue Decomposition (EVD), the natural frequencies and corresponding damping ratios are evaluated as shown in Table 1.

Table 1. Numerical natural frequencies and damping ratios of the 6-DoF storey building system

| Mode Order | 1 | 2 | 3 | 4 | 5 | 6 |
|---|---|---|---|---|---|---|
| $f_i$ (Hz) | 0.0767 | 0.2257 | 0.3616 | 0.4765 | 0.5637 | 0.6181 |
| $\zeta_i$ | 0.0208 | 0.0071 | 0.0045 | 0.0035 | 0.0030 | 0.0028 |

The identification of the time-delay DMDc and the corresponding pseudo-stabilization diagrams are introduced below.

**Step 1: Data collection**

The system response was sampled with the sampling frequency of $f_s = 1/\Delta t = 4$ Hz over a 1000 s observation period. The external excitation is an impulse force applied to the top floor of the building. The resulting impulse responses of all six floors are illustrated in Fig.1 (b).

**Step 2: State matrices**

The state vector of the building is formulated by time-delay embedding of the measured displacement responses:

$$\mathbf{X} = \begin{bmatrix} \mathbf{x}_0(1) & \cdots & \mathbf{x}_0(K-1) \\ \vdots & \ddots & \vdots \\ \mathbf{x}_\tau(1) & \cdots & \mathbf{x}_\tau(K-1) \end{bmatrix} \text{ and } \mathbf{X}' = \begin{bmatrix} \mathbf{x}_0(2) & \cdots & \mathbf{x}_0(K) \\ \vdots & \ddots & \vdots \\ \mathbf{x}_\tau(2) & \cdots & \mathbf{x}_\tau(K) \end{bmatrix} \tag{42}$$

where $\mathbf{x}_\tau(k) = \mathbf{y}(k-\tau)$ with $\mathbf{y}(k) = [y_1(k), \ldots, y_6(k)]^\mathrm{T}$ for $k = 1, \ldots, K \in \mathbb{Z}^+$ and the delay order under an impulse excitation is $\tau \geq 2$ without measurement noise. The input matrix is defined as

$$\mathbf{\Gamma} = \begin{bmatrix} \mathbf{\gamma}_0(1) & \cdots & \mathbf{\gamma}_0(K-1) \\ \vdots & \ddots & \vdots \\ \mathbf{\gamma}_\tau(1) & \cdots & \mathbf{\gamma}_\tau(K-1) \end{bmatrix} \tag{43}$$

where $\mathbf{\gamma}_\tau(k) = \mathbf{u}(k-\tau)$ with $\mathbf{u}(k) = [\underbrace{u(k), 0, \ldots, 0}_{\text{6 elements}}]^\mathrm{T}$.

## Step 3: Time-delay DMDc

The state-space representation of the building system is obtained as $\mathbf{X}' = \mathbf{A}_{\text{DMDc}}\mathbf{X} + \mathbf{B}_{\text{DMDc}}\boldsymbol{\Gamma}$, where the associated eigenvalues and eigenvectors are then identified using the time-delay DMDc approach described in Section 2.2. Here, the truncation orders for input matrix $p$ and output matrix $r$ are determined based on the singular entropy increment criteria:

$$\Delta E_i = -\frac{\lambda_i}{\sum_{i=1}^{l}\lambda_i} \ln\left(\frac{\lambda_i}{\sum_{i=1}^{l}\lambda_i}\right) \tag{44}$$

Where $l$ is the total number of singular values. The singular entropy increment $\Delta E_i$ stabilizes beyond a specific model rank, consistent with the inherent spectral decay of the system eigenvalues. The rank $r$, $p$ are therefore selected at the point where $\Delta E_i$ and its first-order variation approach zero. For simplicity, the truncation rank $p$ is used as an example. Both singular entropy increment and its variation converge to zero from 14 as shown in Fig. 2. Thus, the truncation rank $p$ is selected as 14.

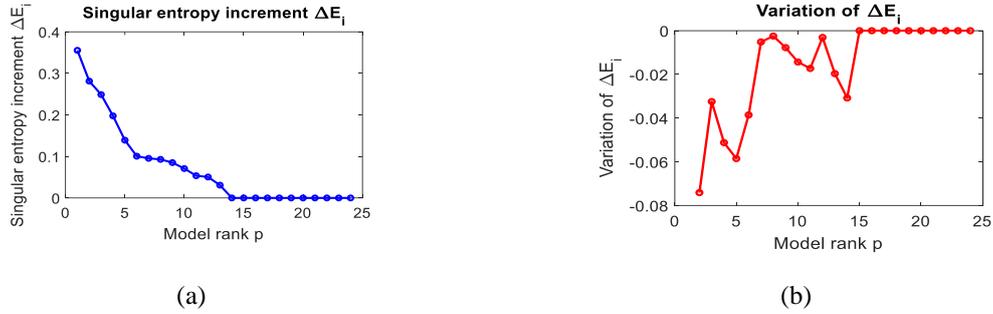

(a)          (b)

Fig. 2 The singular entropy increment and its first-order variation

## Step 4: Modal parameters

Using the time-delay DMDc algorithm, the eigenvalues and eigenvectors of the state transition matrix $\mathbf{A}_{\text{DMDc}}$ are identified, from which the natural frequencies and damping ratios are derived and displayed in frequency and damping pseudo-stabilization diagrams in Fig.3 (a) and (b). To validate the performance of the proposed time-delay DMDc approach, the pLSCF method is conducted using the same experimental data. The stabilization diagrams of both natural frequencies and damping ratios obtained from pLSCF, where the model order is the same as the time-delay embedding order, are presented in Fig. 3 (c) and (d). The colored dashed lines indicate the theoretical natural frequencies and damping ratios.

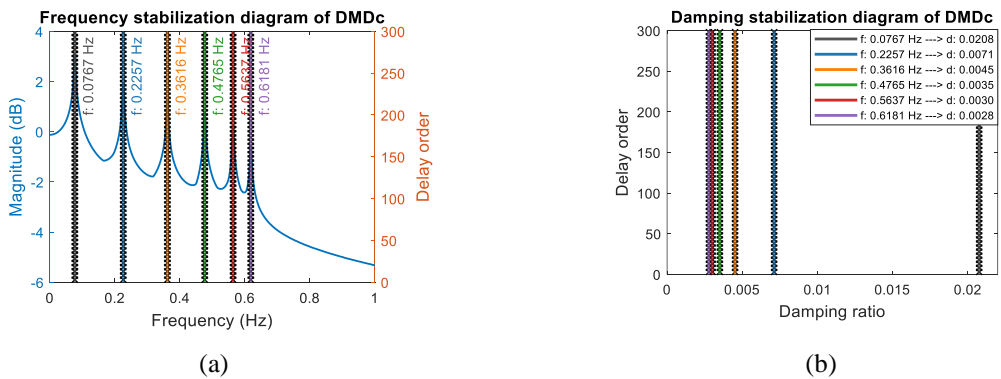

(a)          (b)

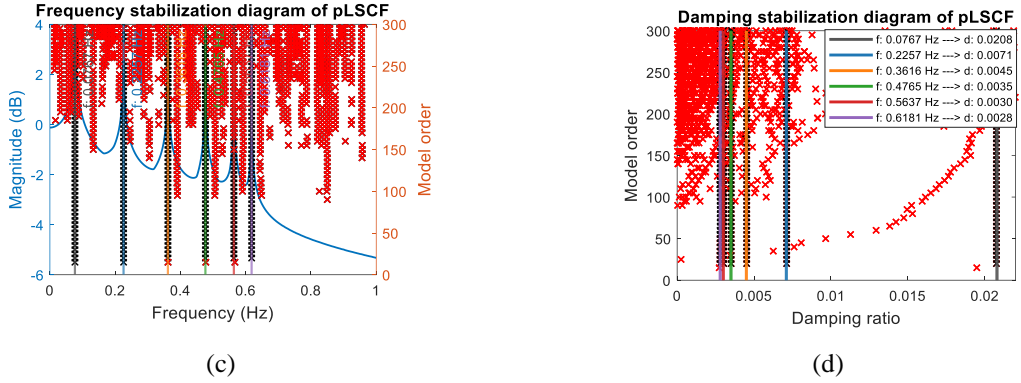

|     |     |
| :-: | :-: |
| (c) | (d) |

Fig. 3. EMA flowchart for the 6-DoF storey building system using time-delay DMDc

As shown in Fig. 3, both time-delay DMDc and pLSCF accurately identify the natural frequencies and damping ratios. However, the pLSCF results contain a large number of numerical modes due to the absence of an intrinsic rank-truncation mechanism as highlighted by red markers. An additional post-processing step is required to reject these numerical roots and improve interpretability. For example, according to the traditional standards in pLSCF-based identification, a mode is categorized as stable only if it satisfies the following relative tolerance thresholds: the variation in natural frequency between successive orders must be below 1% (or 2% depending on noise levels), and the variation in the damping ratio must be below 5%[32]. After post-processing, these accepted physical modes are highlighted by black markers in Fig. 3. In contrast, time-delay DMDc incorporates SVD-based rank truncation of the snapshot matrix, enabling the system dynamics to be captured within a low-dimensional subspace spanned by the dominant singular vectors. This formulation effectively suppresses spurious modes and facilitates efficient analysis of high-dimensional data.

To assess the generalizability of the time-delay DMDc framework under non-impulsive conditions, the initial excitation is substituted with a logarithmic chirp signal ranging from 0.01 Hz to 2 Hz, as illustrated in Fig. 4(a). Unlike impulses, the time-varying nature of the chirp signal provides a rigorous test for the algorithm's ability to decouple intrinsic system dynamics from continuous external forcing. The time-delay DMDc identification procedure, as illustrated above, is then repeated for delay orders ranging from 2 to 53, with a step size of 1, based on the theoretical upper bound of the delay order defined in Eq. (32). For each delay order, the truncation ranks $p$ and $r$ are determined using the singular entropy increment criterion. The observed consistency between the identified modal parameters and their theoretical solutions, as demonstrated in Fig. 4(b) and (c), indicates that the time-delay DMDc effectively filters out the transient complexities of the chirp excitation, achieving accurate identification of the system's modal parameters.

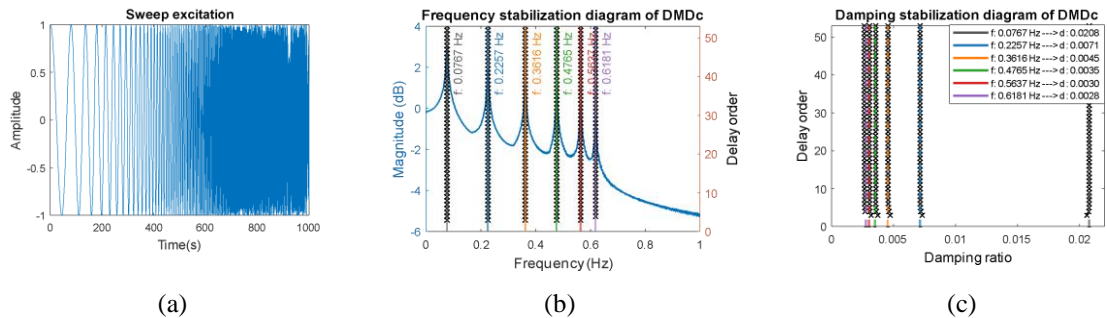

|     |     |     |
| :-: | :-: | :-: |
| (a) | (b) | (c) |

Fig. 4. Identification results under logarithmic chirp excitation: (a) input sweep signal; (b) pseudo-stabilization diagram of natural frequencies; (c) pseudo-stabilization diagram of damping ratios.

## 3.2 EMA with measurement noises

Next, to evaluate the robustness of the time-delay DMDc framework against measurement noise, a sensitivity analysis is performed using response signals corrupted with SNRs ranging from 10 to 40 dB in steps of 3 dB. According to the Eq. (36), the delay order $\tau$ is varied between 53 and 300 with a step size of 5. The identification performance is benchmarked against the pLSCF algorithm, focusing on the 1st and 6th modes to represent the system's low- and high-frequency dynamics. Fig. 5 illustrates the averaged modal identification results with 95% confidence intervals (CIs) for all SNRs, providing a statistical measurement of the estimation's variance and reliability.

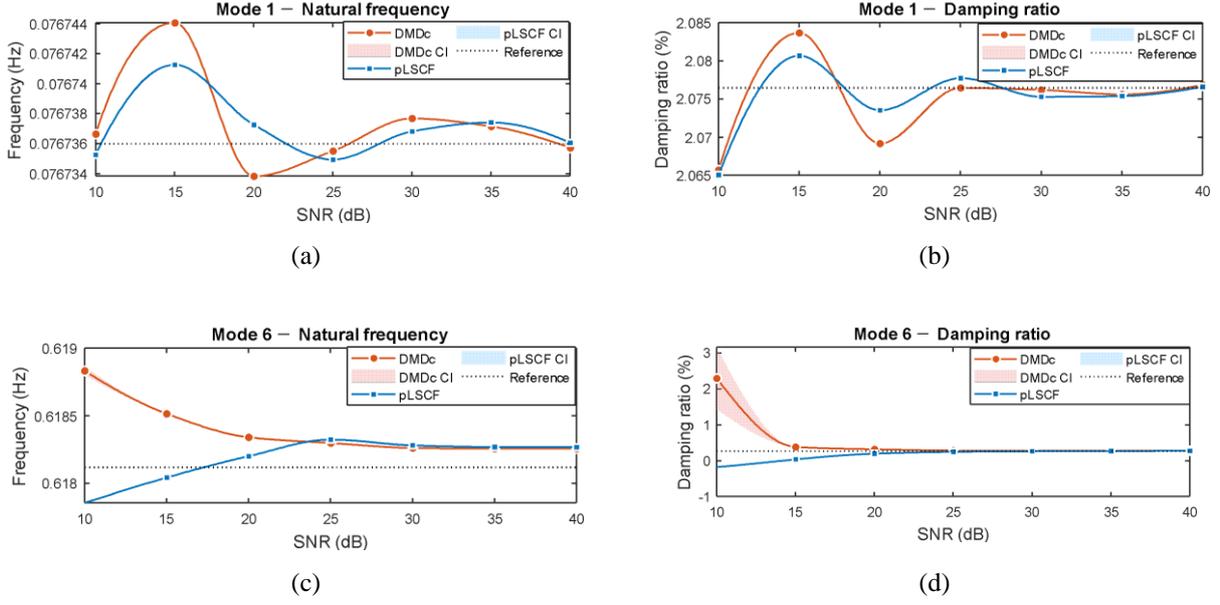

Fig. 5. Robustness comparison of time-delay DMDc and pLSCF under varying SNRs

The 95% CI width is computed based on the standard deviation of repeated simulations for each mode and SNR level. For natural frequency, it is given by:

$$\text{Freq CI}_i = 1.96 \times \frac{1}{n_{\text{modes}} \cdot n_{\text{SNR}}} \sum_{k=1}^{n_{\text{modes}}} \sum_{j=1}^{n_{\text{SNR}}} \sigma_{f,k,j}^{(i)} \tag{45}$$

Where $\sigma_{f,k,j}^{(i)}$ denotes the standard deviation of the estimated natural frequency of mode $k$ at SNR level $j$ using method $i$. Similarly, 95% CIs width of damping ratios are calculated.

The time-delay DMDc method achieves accuracy comparable to the pLSCF method under high SNR conditions. However, time-delay DMDc demonstrated higher noise robustness compared to pLSCF for higher-order modes under low SNR conditions. In contrast to the parametric, model-dependent pLSCF, the non-parametric, SVD-based time-delay DMDc extracts dominant dynamics directly from data, inherently reducing the impact of measurement noise and model mismatch. This leads to practical advantage, for example, severe noise conditions (SNR down to 10 dB), pLSCF fails to identify the 6th mode, while time-delay DMDc maintains stable frequency estimates, demonstrating its stronger capability to handle uncertainty, though certain errors persist in damping ratio identification.

Specifically, using a SNR of 20 as a representative example, the effect of measurement noise on time-delay DMDc performance with increasing delay order is systematically analyzed using pseudo-stabilization diagrams of natural frequencies and damping ratios, with comparisons to the corresponding pLSCF stabilization diagrams shown in Figs. 6 and 7. The results demonstrate that time-delay DMDc yields natural frequency estimates that remain consistent with theoretical values and comparable to those obtained by pLSCF, while damping estimation is more sensitive to noise, particularly for higher-order modes. However, increasing the delay and model orders leads to a progressive stabilization of the damping ratio estimates for both methods, indicating that the introduction of time-delay effectively improves robustness to measurement noise, as evidenced in the appendix A.

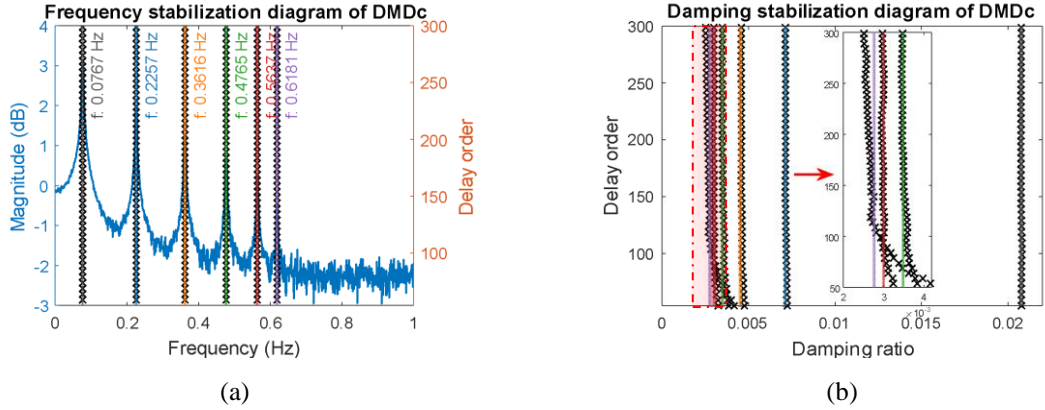

(a)                                                  (b)

Fig. 6. Pseudo-stabilization diagrams computed by time-delay DMDc for 6-DoF storey building system with measurement noise SNR=20

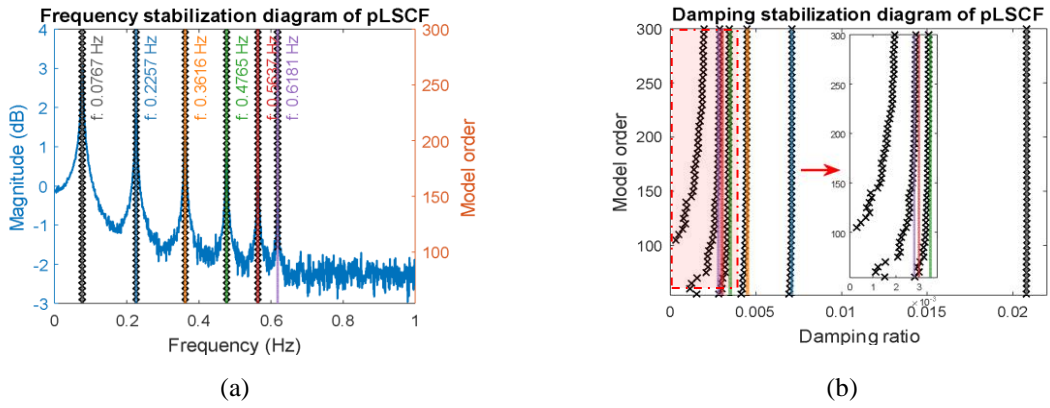

(a)                                                  (b)

Fig. 7. Stabilization diagrams computed pLSCF for 6-DoF storey building system with measurement noise

A more objective comparison of bias and uncertainty between time-delay DMDc and pLSCF is conducted using ensemble statistics of modal estimates over varying delay and model orders. Fig. 8 presents a histogram-based comparison of the natural frequencies and damping ratios identified by time-delay DMDc and pLSCF across all delay and model orders, with results from time-delay DMDc shown in subplots (a)–(d) and those from pLSCF in subplots (e)–(h). The statistical distributions of the estimates are fitted with Gaussian probability density functions, with the mean values indicated by dashed black lines together with the reference values. For clarity, only the 1st and 6th natural frequencies and damping ratios are shown. Table 2 summarizes the corresponding mean estimates and their errors relative to the analytical solutions for both methods over all delay and model orders. From Fig. 8 and Table 2, it can be observed that both methods accurately identify natural frequencies, with pLSCF showing marginally lower variance, and damping ratio estimation proves more challenging and more sensitive to measurement noise, with a good agreement in the zoomed views of Figs. 6 and 7. Notably, time-delay DMDc provides unbiased damping estimates even for higher-order modes, whereas pLSCF exhibits a more pronounced systematic bias despite its lower variance. Consequently, time-delay DMDc demonstrates superior robustness than pLSCF, especially for damping ratio identification in the presence of measurement noise.

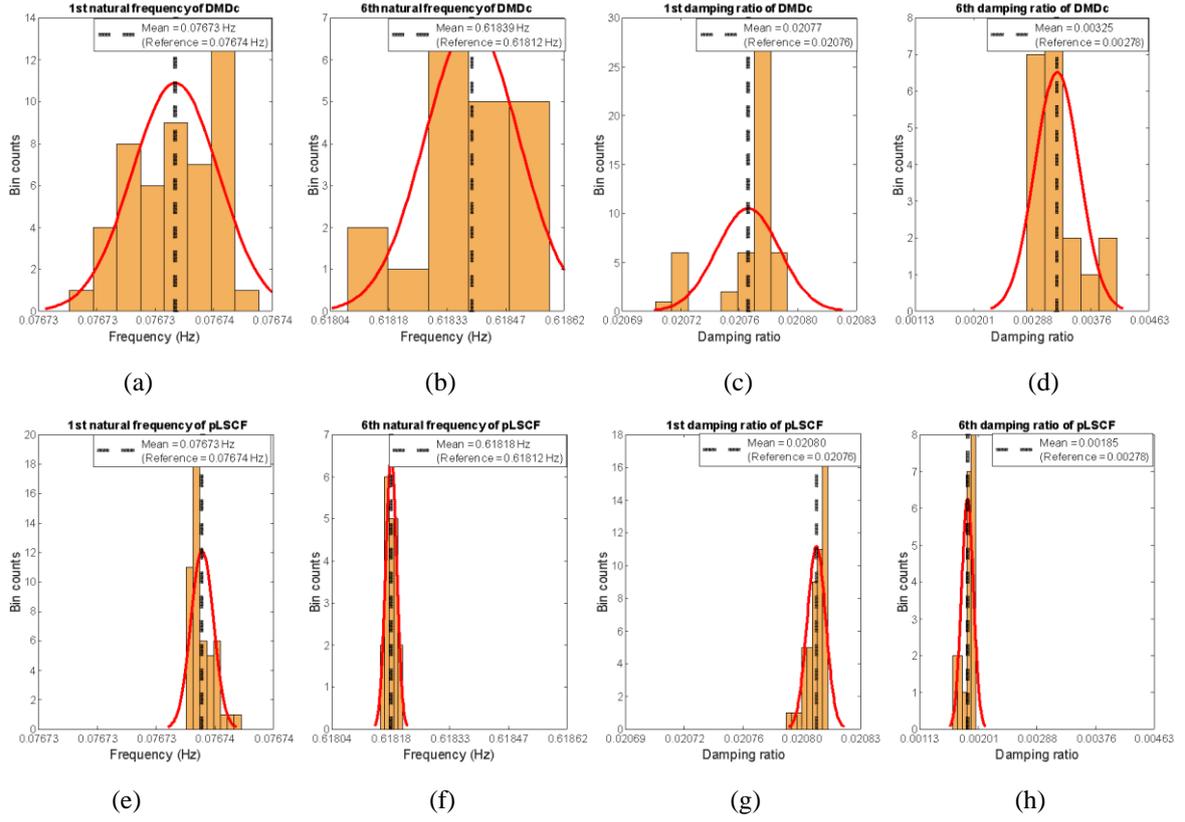

Fig. 8. Uncertainty quantification of the 1st and 6th modal parameters via time-delay DMDc and pLSCF

Table 2. Comparison of the average identified modal parameters by time-delay DMDc and pLSCF with analytical solutions

| Mode order | $f_i$(Hz) | | | | $\zeta_i$ | | | |
|---|---|---|---|---|---|---|---|---|
| | pLSCF | DMDc | pLSCF Error(%) | DMDc Error(%) | pLSCF | DMDc | pLSCF Error(%) | DMDc Error(%) |
| 1 | 0.0767 | 0.0767 | 0.0000 | 0.0000 | 0.0208 | 0.0208 | 0.0000 | 0.0000 |
| 2 | 0.2258 | 0.2257 | 0.0443 | 0.0000 | 0.0071 | 0.0072 | 0.0000 | 1.4100 |
| 3 | 0.3616 | 0.3617 | 0.0000 | 0.0277 | 0.0044 | 0.0045 | 2.2200 | 0.0000 |
| 4 | 0.4766 | 0.4765 | 0.0210 | 0.0000 | 0.0033 | 0.0035 | 5.7100 | 0.0000 |
| 5 | 0.5637 | 0.5637 | 0.0000 | 0.0000 | 0.0026 | 0.0031 | 13.330 | 3.3300 |
| 6 | 0.6182 | 0.6184 | 0.0162 | 0.0485 | 0.00185 | 0.0030 | 33.930 | 7.1400 |

To improve computational efficiency and stabilize damping ratio estimation at lower delay orders, the system response is downsampled according to the Nyquist–Shannon sampling criterion. By suppressing high-frequency measurement noise outside the modal band, downsampling effectively enhances the SNR in the frequency range of interest (see Appendix A), thereby enabling more stable damping estimation with reduced delay orders. In this csae, the sampling frequency is reduced from 4 Hz to 1.4 Hz, and the resulting identification results shown in Fig. 9 demonstrate that damping ratio estimates stabilize at significantly lower delay orders compared with the original sampling case (Fig. 6 (b)), indicating the effectiveness of downsampling in improving robustness and efficiency in time-delay DMDc.

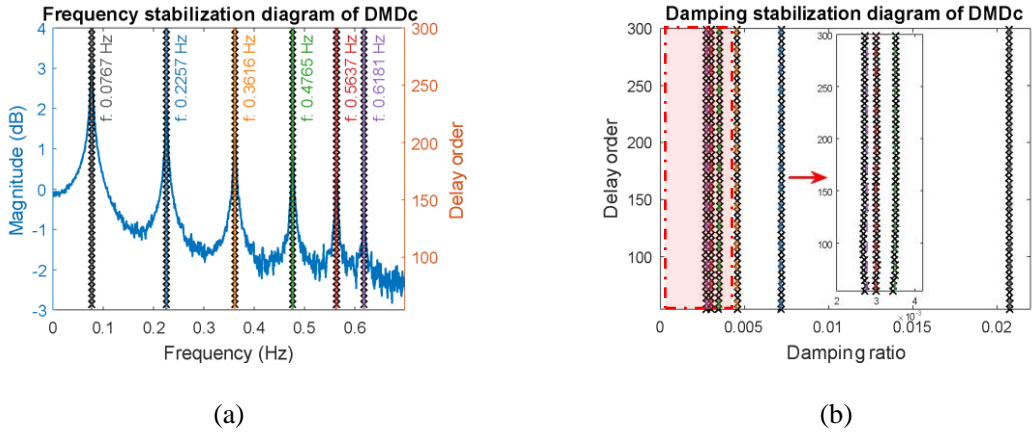

(a)          (b)

Fig. 9. Pseudo-stabilization diagrams computed by time-delay DMDc for 6-DoF storey building system with measurement noise after downsampling

The accuracy of the mode shapes extracted by the time-delay DMDc and pLSCF is also discussed. The time-delay DMDc mode shapes have been computed by Eq. (32). For comparison, the numerical solutions are shown in Fig. 10. One method for quantifying differences between mode shapes is the Modal Assurance Criterion (MAC) gives a scalar constant relating the degree of consistency between one modal shapes and another reference modal shapes as defined by Eq. (46):

$$\text{MAC}(\phi_i, \psi_i) = \frac{|\phi_i^H \psi_i|^2}{(\phi_i^H \phi_i)(\psi_i^H \psi_i)} \tag{46}$$

$\phi_i$ is the $i^{th}$ estimated mode shape vector, $\psi_i$ is $i^{th}$ reference mode shape vector. The calculated scalar values range between 0 and 1, where 1 indicates that the two mode shapes are identical and 0 indicates that they are completely uncorrelated. The MAC values between the time-delay DMDc modal shapes and the analytic modal shapes are shown in Fig. 11 (a). The MAC values between the pLSCF modal shapes and the analytic modal shapes are shown in Fig. 11 (b). Clearly, the time-delay DMDc mode shapes are in close agreement with the numerical solutions and the pLSCF mode shapes.

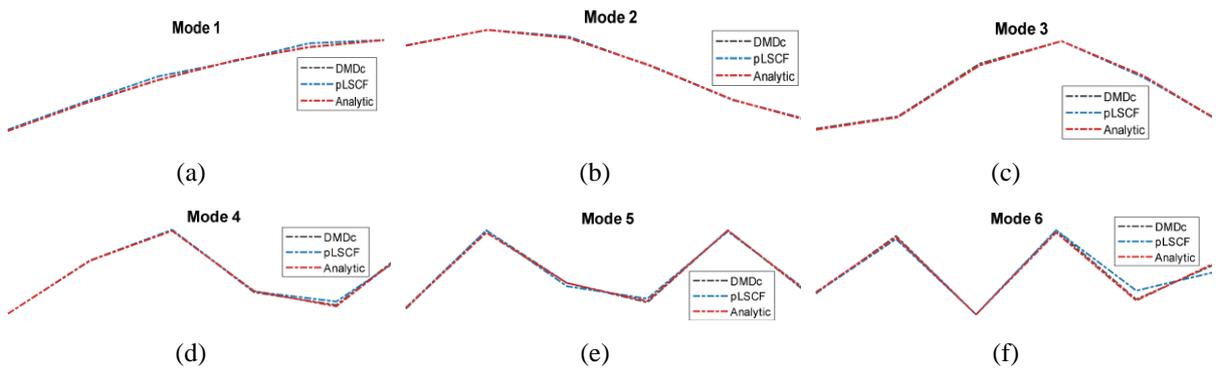

Fig. 10. Mode shapes: time-delay DMDc vs. pLSCF vs. analytic mode shapes

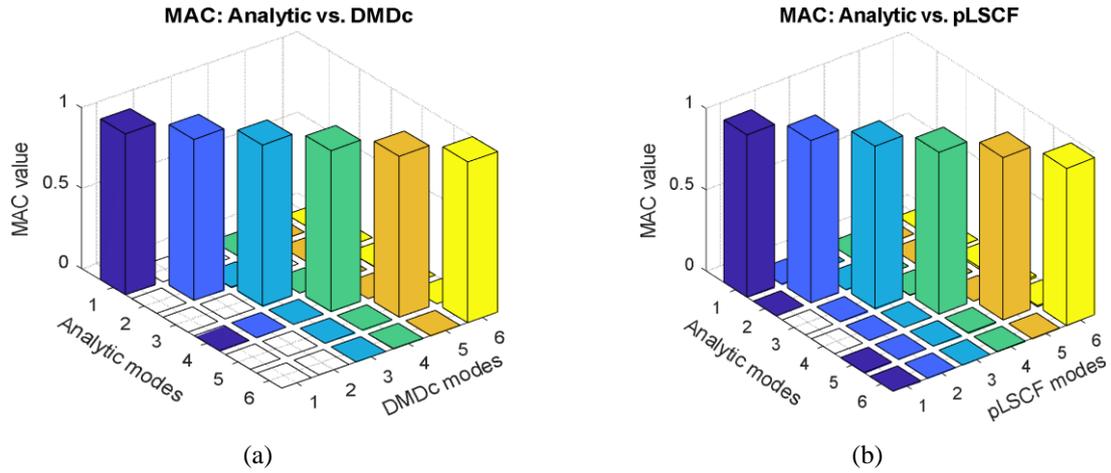

Fig. 11. MAC comparison: Analytic mode shapes vs. time-delay DMDc vs. pLSCF

## 4 Experimental study on a cantilever beam

This section investigates the effectiveness of time-delay DMDc in estimating the modal parameters of the cantilever beams under impact excitation as well as its capability to handle high-dimensional (e.g., full-field measurement) data from digital cameras.

### 4.1 Modal parameter identification using time response acquired from the accelerometers

The steel cantilever beam has a length of 420 mm, a width of 2 mm, and a height of 30 mm. The following describes the procedure for identifying modal parameters of cantilever beams, consistent with the previously defined 6-DoF time-delay DMDc EMA framework.

Step 1, acceleration responses were acquired from three accelerometers attached at distances of 140 mm, 245 mm, and 345 mm from the fixed end of the beam under impact excitation, with a sampling frequency of 8192 Hz. The experimental setup, the corresponding input and output time histories are detailed in Fig. 12.

Step 2, the measured acceleration were processed within the time-delay DMDc framework to construct an augmented state representation. The minimum delay order was determined according to Eq. (36). Then, the delay order range was set from 250 to 300 with a step size of 1.

Step 3, the state-space representation of the beam is constructed, while the truncation ranks $p$ and $r$ were selected using the singular entropy increment criterion applied to $\Omega$ and $\mathbf{X}'$.

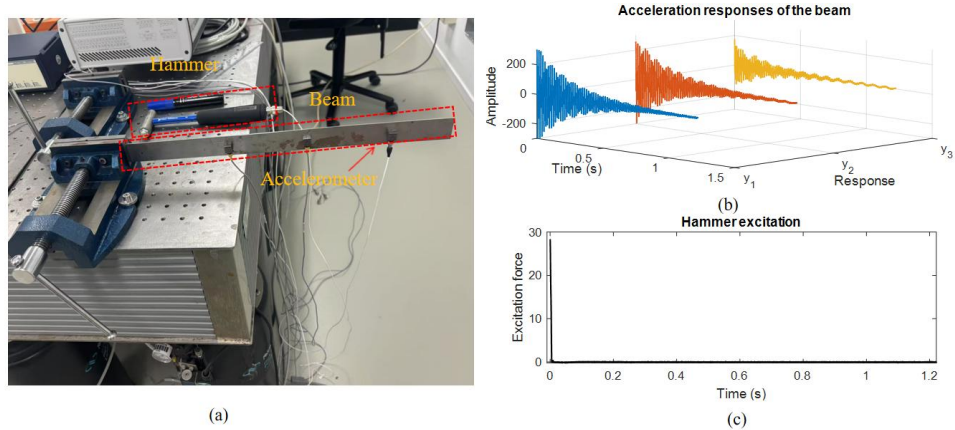

Fig. 12. Experimental setup for measuring vibrations on a cantilever beam: (a)Beam vibration test setup, (b)The acceleration signals and (c) Hammer excitation

Step 4, the robustness of the time-delay DMDc was validated by the resulting pseudo-stabilization diagrams about the natural frequencies and damping ratios, as shown in Fig. 13. These results are then compared with the modal parameter stabilization diagrams identified by the pLSCF at the model order that aligns with the delay order, as illustrated in Fig. 14. The dashed lines represent the mean values of the identified modal parameters across all delay and model orders. A quantitative comparison between the two methods is presented in Table 3, which summarizes the mean natural frequencies and damping ratios identified by time-delay DMDc and pLSCF, together with their relative differences.

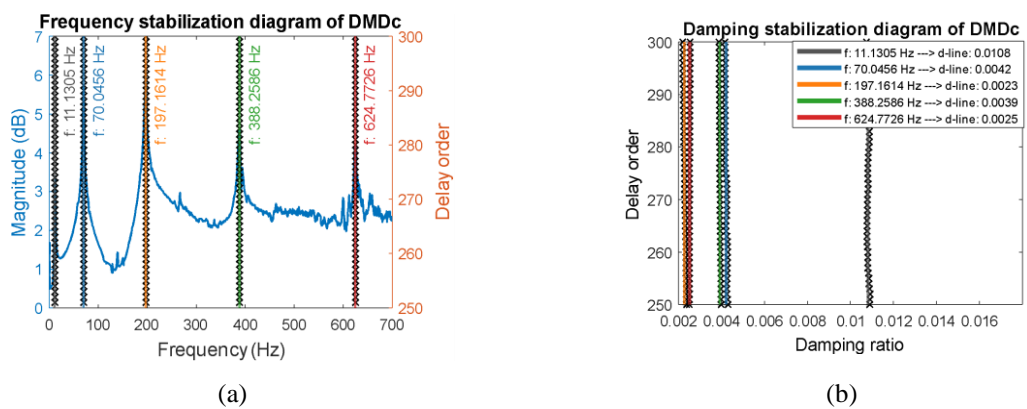

Fig. 13. Pseudo-stabilization diagrams of natural frequencies from time-delay DMDc

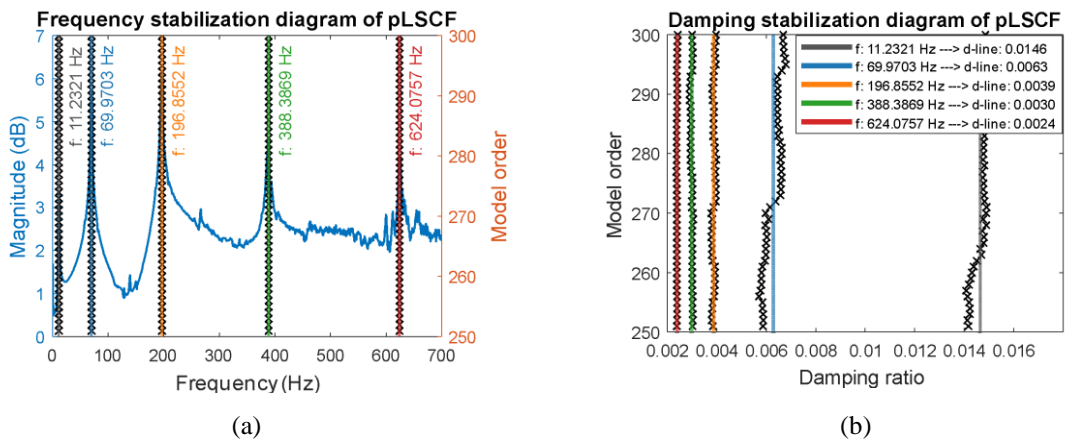

Fig. 14. Stabilization diagrams of natural frequencies from pLSCF

Table 3. Comparison of the average identified modal parameters by time-delay DMDc and pLSCF for the cantilever beam

| Mode order | $f_i$(Hz) | | | $\zeta_i$ | | |
|---|---|---|---|---|---|---|
| | pLSCF | DMDc | Relative Error(%) | pLSCF | DMDc | Relative Error(%) |
| 1 | 11.2321 | 11.1305 | 0.9000 | 0.0146 | 0.0108 | 26.0300 |
| 2 | 69.9703 | 70.0456 | 0.1100 | 0.0063 | 0.0042 | 33.3300 |
| 3 | 196.8552 | 197.1614 | 0.1600 | 0.0039 | 0.0023 | 41.0300 |
| 4 | 388.3869 | 388.2586 | 0.0300 | 0.0030 | 0.0039 | 30.0000 |
| 5 | 392.2488 | 390.5849 | 0.4200 | 0.0072 | 0.0059 | 18.0600 |
| 6 | 624.0757 | 624.7726 | 0.1100 | 0.0024 | 0.0025 | 4.1700 |

The results indicate that both methods yield stable stabilization diagrams and nearly identical estimates of natural frequencies across all modes. In contrast, more pronounced discrepancies are observed in the damping ratio estimates, reflecting the higher sensitivity of damping identification to measurement noise due to its small magnitude.

To evaluate the stability of damping ratio estimation, boxplots were constructed as shown in Fig. 15, where each box represents the statistical distribution of the identified damping ratios for a given mode over the considered range of model and delay orders. The central line denotes the median value, while the box spans the interquartile range (Q1–Q3), reflecting the variability of the estimates. The whiskers indicate the range within 1.5 times the interquartile range, capturing the extent of reasonable fluctuations. A narrower box implies higher robustness with measurement noise. As illustrated in Fig. 15, time-delay DMDc provides consistently stable damping ratio estimates for all identified modes. By comparison, the first and second modes identified by pLSCF exhibit relatively larger interquartile ranges, indicating a higher sensitivity to measurement noise compared with time-delay DMDc.

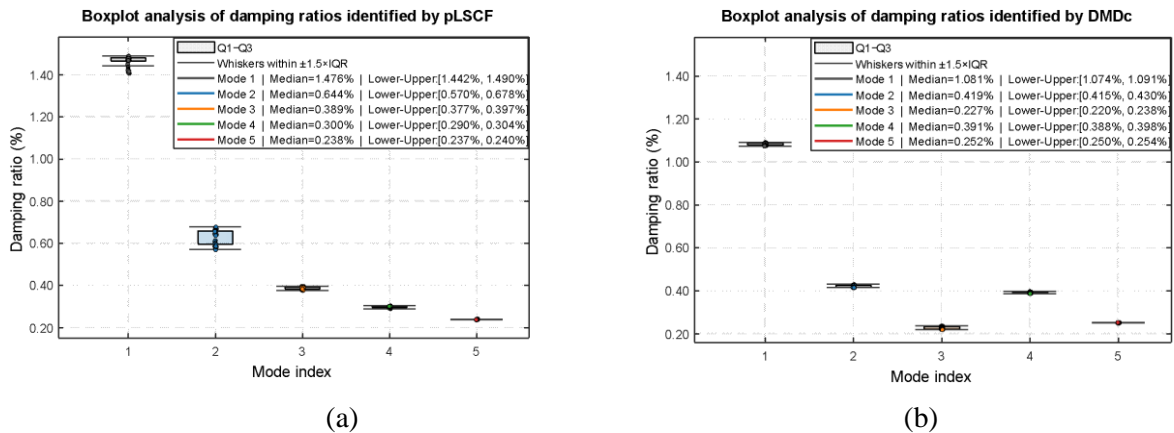

(a)  (b)

Fig.15. Boxplot analysis of damping ratios identified by pLSCF and time-delay DMDc

### 4.2 Modal parameter identification for full-field measurements using digital camera

Further validation of the feasibility of time-delay DMDc in handling high-dimensional data is conducted through the extraction of modal parameters from image-based large-scale data obtained through a high-speed camera (i.e.,) under hammer impact.

Step 1, the experimental setup follows the configuration shown in Fig. 16 (a). The steel cantilever beam has dimensions of 420 mm in length, 2 mm in width, and 30 mm in height. The beam is initially excited by a hammer impact, after which a high-speed CMOS camera, illustrated in Fig. 16 (b), records the vibration response after conducting camera calibration. The motion trajectories of 1207 pixels along the beam edge are extracted using CoTracker[28] and treated as vibration responses for the subsequent time-delay DMDc modeling. The recorded data consist of 601 sampling points with a sampling frequency of 120 Hz. For illustrative purposes, the motion of the first six representative pixels are shown in Fig. 16 (c). According to Proposition 1, the hammer impact is idealized as a unit impulse excitation, as depicted in Fig. 16 (d), and is employed for time-delay DMDc modeling of the high-dimensional image-based data. Fig. 17 illustrates the motion trajectories of all identified edge points under a zoomed-in region.

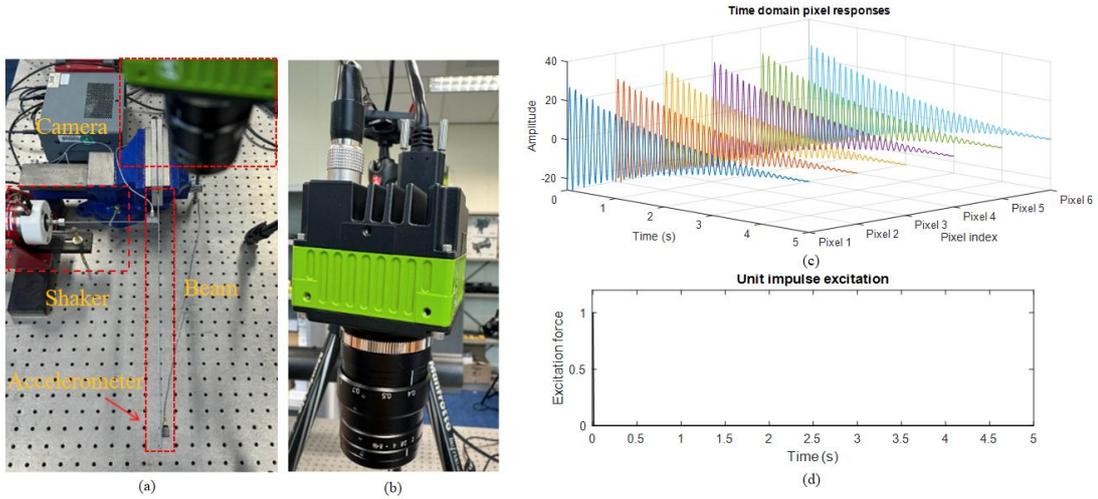

Fig. 16. Experimental setup for measuring vibrations on a cantilever beam: (a) overall experimental setup; (b) high-speed camera with a high-resolution lens (adapted and modified from Ref.[28] under CC BY 4.0) (c)The responses and (d) Hammer excitation

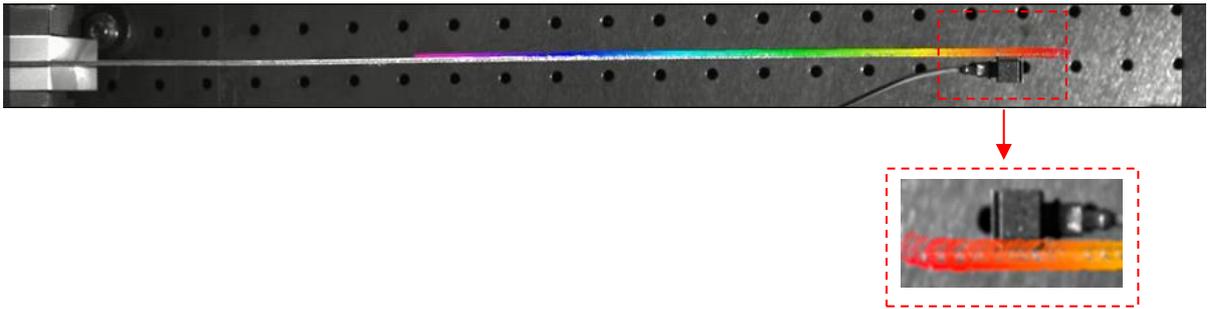

Fig. 17. The trajectory plot of the edge points with a zoomed-in view of the region

Step 2, the delay order can be obtained using Eq. (36), set to 200 to 250, with a step size of 1. The time-delay embedded snapshot matrix was constructed for time-delay DMDc modeling, with dimensions of (number of delays × number of pixels) × (number of sampling points- number of delays).

Step 3, The augmented state-space model of the beam was constructed and the truncation ranks for input matrix $p$ and output matrix $r$ are determined based on the singular entropy increment criteria.

Step 4, the pseudo-stabilization diagrams of the natural frequencies and damping ratios obtained from time-delay DMDc are presented in Fig. 18. The corresponding mode shapes are illustrated in Fig. 19. The time-delay DMDc demonstrates robust identification of the first two modal parameters including natural frequencies (8.5859 Hz and 54.3992 Hz), damping ratios, and mode shapes in the y-direction, while the first order natural frequency in the *x*-direction is 16.8589 Hz, whereas the application of the pLSCF algorithm is restricted by computational limitations when dealing with high-dimensional systems.

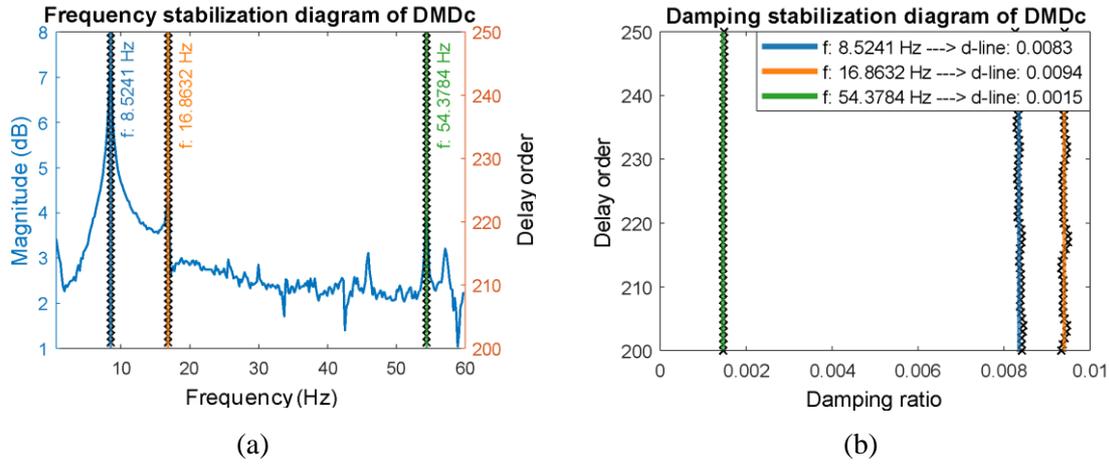

Fig. 18. Pseudo-stabilization diagrams computed by time-delay DMDc for the cantilever beam

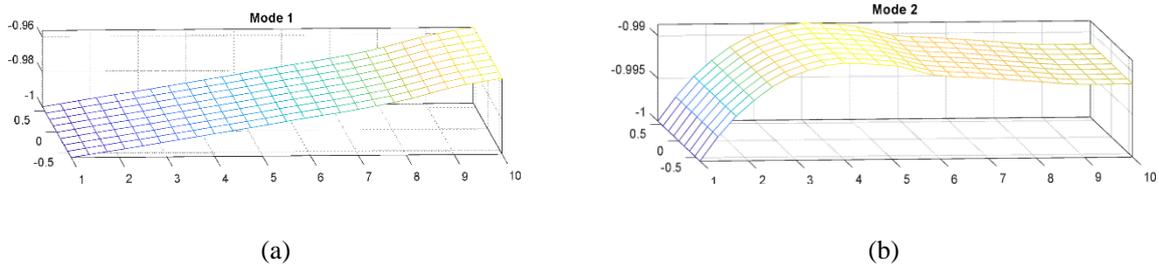

Fig. 19. Mode shapes

## 5  Discussion

In this study, the theoretical connection between pLSCF and time-delay DMDc within the framework of EMA was rigorously established by reformulating both methods under a unified ARX-based state-space representation. This analysis clarifies the physical interpretability of time-delay DMDc, originally developed in fluid mechanics, and provides a theoretical foundation for extending it to structural modal analysis, an aspect that has remained insufficiently understood in existing literature. Building upon this theoretical framework, a time-delay DMDc formulation was developed to address a critical limitation of conventional EMA techniques: the difficulty of performing robust modal identification in high-dimensional measurement scenarios. By incorporating delay-embedding states, the proposed method significantly improves noise robustness and enables stable identification of damping ratios, outperforming classical approaches such as pLSCF in noisy environments while maintaining comparable accuracy in natural frequency and mode shape estimation for low-dimensional systems.

Comprehensive numerical evaluations using a 6-DoF building model demonstrate that time-delay DMDc reliably extracts modal parameters even under severe noise contamination (SNR as low as 10 dB), and that downsampling combined with an appropriate delay order effectively ensures stable and accurate damping estimates at lower delay order. Further experimental validation on cantilever beams, using both accelerometer measurements and high-dimensional video-based motion tracking, confirms that the proposed approach is capable of processing large-scale measurement data, where pLSCF becomes computationally impractical.

Overall, this work establishes time-delay DMDc as a powerful and generalizable paradigm for structural EMA, particularly for high-dimensional sensing technologies enabled by vision-based or dense sensor arrays. Future work will focus on real-world large civil structures such as bridges and buildings, as well as extensions toward operational modal analysis to fully exploit the potential of high-dimensional structural monitoring environments.

# 7. Appendix A

Considering the system described by Eq. (9), the state vector $\mathbf{x}_{\tau_a,k} \in \mathbb{R}^n$ corrupted by additive Gaussian white noise $\mathbf{e}_{\tau_a,k} \in \mathbb{R}^n$

$$\tilde{\mathbf{x}}_{\tau_a,k} = \mathbf{x}_{\tau_a,k} + \mathbf{e}_{\tau_a,k}, \mathbf{e}_{\tau_a,k} \sim \mathbb{N}(0,\sigma^2 \mathbf{I}_n) \tag{A1.1}$$

To enhance robustness, time-delay embedding is performed by stacking $\tau$ consecutive state vectors into an augmented state:

$$\underset{n(\tau_a+1)\times 1}{\tilde{\mathbf{x}}_k} = \begin{bmatrix} \tilde{\mathbf{x}}_{0,k} \\ \vdots \\ \tilde{\mathbf{x}}_{\tau_a,k} \end{bmatrix} = \overline{\mathbf{x}}_k + \mathbf{v}_k, \quad \underset{n(\tau_a+1)\times 1}{\mathbf{v}_k} = \begin{bmatrix} \mathbf{e}_{0,k} \\ \vdots \\ \mathbf{e}_{\tau_a,k} \end{bmatrix} \tag{A1.2}$$

The measurement noise vectors $\mathbf{v}_k$ is zero-mean Gaussian with covariance $\mathbb{E}[\mathbf{v}_k \mathbf{v}_k^\top] = \sigma^2 \mathbf{I}_{(\tau_a+1)n}$. The state matrix $\tilde{\mathbf{X}}$ formed by concatenating error-corrupted state vectors $\{\tilde{\mathbf{x}}_k\}_{k=1}^{K-1}$, performs SVD of $\tilde{\mathbf{X}}$ whose first left singular vector $\tilde{\mathbf{u}}$ represents the dominant left singular vector. Note that the obtained $\tilde{\mathbf{u}}$ is an error-corrupted estimate of the true singular vector.

$$\tilde{\mathbf{u}} \approx \frac{1}{\sqrt{\tau_a}}[\mathbf{u}_0;\ldots;\mathbf{u}_0] + O\left(\frac{\sigma}{\sqrt{\tau_a}}\right) \tag{A1.3}$$

Where $\mathbf{u}_0 \in \mathbb{R}^n$ is the signal-dominant direction. The projection error $\epsilon$ of the $\mathbf{v}_k$ to $\tilde{\mathbf{u}}$ captures the instantaneous deviation in the time-delay DMDc mode direction due to measurement noise.

$$\epsilon = \tilde{\mathbf{u}}^\mathrm{T} \mathbf{v}_k \approx \underbrace{\left(\frac{1}{\tau_a}\sum_{i=1}^{\tau_a} \mathbf{e}_{i,k}^\mathrm{T} \mathbf{u}_0\right)}_{\text{Dominant term}} + O\left(\frac{\sigma^2}{\tau_a}\right) \tag{A1.4}$$

The variance and the standard deviation of the projection error $\epsilon$ are:

$$\mathrm{Var}\left(\frac{1}{\tau_a}\sum_{i=1}^{\tau_a} e_{i,k}^\mathrm{T} \tilde{\mathbf{u}}_0\right) = \frac{1}{\tau_a} \cdot \sigma^2 \| \mathbf{u}_0 \|^2 = \frac{\sigma^2}{\tau_a} \tag{A1.5}$$

$$\mathrm{Std}(\epsilon) = \frac{\sigma}{\sqrt{\tau_a}} \tag{A1.6}$$

Time-delay DMDc computes eigenvalues by perturbed by $\delta u_i$ from projected error data. The perturbation scales as:

$$\delta u_i \sim \epsilon \quad \Rightarrow \quad |\delta u_i| \sim \frac{\sigma}{\sqrt{\tau_a}} \tag{A1.7}$$

The corresponding continuous-time eigenvalue changes approximately as:

$$s_i + \delta s_i \approx f_s \cdot \log(u_i + \delta u_i) \tag{A1.8}$$

Applying a first-order Taylor approximation:

$$\delta s_i \approx \frac{f_s}{u_i} \cdot \delta u_i \tag{A1.9}$$

Taking magnitudes:

$$|\delta s_i| \approx \frac{f_s}{|u_i|} \cdot |\delta u_i| \sim \frac{f_s \sigma}{\sqrt{\tau_a}} \tag{A1.10}$$

Since damping ratio is computed from $s_i$, the perturbation also scales similarly:

$$|\delta \zeta_i| \sim \frac{f_s \sigma}{\sqrt{\tau_a}} \tag{A1.11}$$

The delay order $\tau_a$ of time-delay DMDc leads to a $\frac{1}{\tau_a}$ reduction in the projection variance of measurement noise, and hence a $\frac{1}{\sqrt{\tau_a}}$ suppression of modal parameter perturbation. Note that, for damping ratio, in lightly damped systems, the estimation of damping ratios is inherently sensitive to measurement noise. This arises because it relies on the real part of the eigenvalues, which is typically orders of magnitude smaller than the imaginary part (associated with natural frequencies). Consequently, even minor deviations in the real component induced by measurement or model noise, which can result in disproportionately large relative noise in the estimated damping ratios.